\documentclass[twocolumn,amsmath,amssymb,floatfix]{snp}
\usepackage{tikz}
\usepackage{graphicx}
\usepackage{dcolumn}
\usepackage{bm}
\newcommand{\bea}{\begin{eqnarray}}
\newcommand{\eea}{\end{eqnarray}}
\newcommand{\be}{\begin{equation}}
\newcommand{\ee}{\end{equation}}

\def\be{\begin{eqnarray}}
\def\ee{\end{eqnarray}}
\def\bd{\begin{displaymath}}
\def\ed{\end{displaymath}}

\def\ga{\gamma}
\def\sl{^-\!\!\!\!}
\def\ADNDT{{At. Data Nucl. Data Tables }}

\def\NP{Nucl. Phys. }
\def\PR{Phys. Rev. }
\def\PRL{Phys. Rev. Lett. }
\def\PL{Phys. Lett. }
\def\jpg{J. Phys. G: Nucl. Part. Phys. }
\def\EPJ{{Eur. Phys. J. }}
\def\etal{{et al.}}

\begin{document}

\title{{\Large Study of reaction and decay using densities from relativistic mean field theory}}

\author{\large G. Gangopadhyay}
\email{ggphy@caluniv.ac.in}
\affiliation{Department of Physics, University of Calcutta, Kolkata - 700009, INDIA}
\begin{abstract}
\leftskip1.0cm
\rightskip1.0cm
Relativistic mean field calculations have been performed to obtain 
nuclear density profile. Microscopic interactions have been folded with the 
calculated densities of finite nuclei to obtain a semi-microscopic potential. 
Life time values for the emission of proton, alpha particles and complex 
clusters have been calculated in the WKB approach assuming a tunneling process 
through the potential barrier. Elastic scattering cross sections have been 
estimated for proton-nucleus scattering in light neutron rich nuclei. 
Low energy proton reactions have been studied and their astrophysical 
implications have been discussed.
The success of the semi-microscopic potentials obtained in the folding 
model with RMF densities in explaining nuclear decays and reactions has been
emphasized.
\end{abstract}\maketitle

\section{Introduction}

Relativistic mean field (RMF) approach has proved to be very successful in
explaining different features of stable and exotic nuclei such as ground state 
binding energy, deformation, radius, excited states, spin-orbit splitting, 
neutron halo, etc\cite{RMF1}. Particularly, the radius and the nuclear density 
are known to be well reproduced in this model. In nuclei 
far away from 
the stability valley, the single particle level structure undergoes certain
changes in which the spin-orbit splitting plays an important role.
Being based on the Dirac Lagrangian density, and thus naturally incorporating 
the spin degree of freedom, RMF is particularly suited to investigate these 
nuclei.

RMF calculations have been found to provide good description of densities, 
particularly at large radii. Processes, such as  proton and alpha decay, low 
energy scattering and reaction, probe this part of the nuclear volume. Thus, a 
good 
description of such processes may be expected if we use microscopic 
nucleon-nucleon (NN) interactions and densities from RMF calculations to 
construct nucleon(nucleus)-nucleus potentials. This will be very 
useful to extend any calculation to areas far from the stability valley, where 
data are scarce and difficult to obtain in near future.
For example, low energy reactions are very important from the astrophysical 
point of view. In astrophysical environments, neutron and proton reactions
are the keys to nucleosynthesis of heavy elements. However, very often, the 
target nuclei are not available in terrestrial laboratories and one needs to 
depend on theoretical inputs. 

The presentation has been arranged in the following manner. In Section 2, the 
method of constructing the potential is outlined. The next section presents a few theoretical 
results for density, followed by selected results of calculation for decays 
involving emission of proton, alpha and complex clusters. We also briefly 
discuss the results of elastic proton scattering in neutron rich nuclei. A few 
low energy proton reactions in $A=60-80$ region have also been studied and 
astrophysical importance of such reactions has been discussed,  Finally, we 
summarize our results.

\section{Method}

There is a large body of work for the topics that
have been presented in this talk. In this section, we restrict ourselves to
the salient points of the method of calculation that have been followed to 
obtain the semi-microscopic potentials used in the present work.

Theoretical density profiles have been extracted from  RMF calculations.
In this approach, nucleons interact via exchange of a number of mesons.
There are different variations of the Lagrangian density as well as a number 
of different parametrizations. In the present work we have employed the FSU 
Gold\cite{prl} Lagrangian density. It  contains, apart from the usual component
describing a system of nucleons interacting via exchange of mesons,
nonlinear terms involving self coupling of the scalar-isoscalar 
and the vector-isoscalar meson, as well as coupling between the
vector-isoscalar meson and the vector-isovector meson.

In the conventional RMF+BCS approach for even-even nuclei, the Euler-Lagrange
equations are solved under the assumptions of classical meson
fields, time reversal symmetry, no-sea contribution, etc. Pairing is introduced
under the BCS approximation. Since accuracy of the nuclear density is very
important in our calculation, we have solved the equations in co-ordinate space.
The strength of the zero range pairing force is taken as 300 MeV-fm for both
protons and neutrons. These values have been chosen to represent a good fit
for the binding energy values. In nuclei containing odd number of neutrons or 
protons, the tagging approximation has been used to specify the level occupied 
by the last odd 
nucleon of either type. We have observed that moderate variations of the 
pairing strength do not influence the results to any great extent.

Effective NN interactions, such as density dependent M3Y 
(DDM3Y)\cite{ddm3y1,ddm3y2,ddm3y3}, or that of Jeukenne, Lejeune and Mahaux 
(JLM)\cite{jlm} may be used to
construct the nucleon (nucleus)-nucleus potential. Both the interactions 
mentioned above have been derived from nuclear matter calculation and have been applied in finite nuclei
with success.

The DDM3Y interaction\cite{ddm3y1,ddm3y2} is obtained from a finite range
energy independent M3Y interaction by adding a zero range energy dependent
pseudopotential and introducing a density dependent factor.
The density dependence may be chosen as exponential\cite{ddm3y1} or be
of the form  $C(1-\beta\rho^{2/3})$\cite{ddm3y2}.
The constants were obtained from nuclear matter calculation\cite{ddm3y3}
as $C=2.07$ and $\beta=1.624$ fm$^2$. We have adopted all the standard
parameters in our calculation without any modification. This
interaction has been employed widely in the study of nucleon nucleus as well
as nucleus nucleus scattering and radioactivity.

In the JLM potential\cite{jlm}, 
finite range of the interaction has been introduced by
including a Gaussian form factor\cite{jlm1}. 
Here, we have adopted the 
global parameters for the interaction and the default normalizations\cite{jlm1}. 

The semi-microscopic nuclear potentials have been obtained by folding the
interactions with the microscopic densities obtained in the RMF 
calculation.
The Coulomb potential has similarly been obtained by folding the Coulomb
interaction with the microscopic proton densities.  
The spin-orbit potential was
chosen from the Scheerbaum prescription\cite{SO}.
The total potential
consists of the nuclear part, the Coulomb potential as well as the centrifugal
potential. We have not included the contribution of isovector (Lane)
potential. However, we expect its effect
to be
small.

\section{Results}

\subsection{Density}

\begin{figure}[hbt]
\vskip -4cm
\resizebox{11cm}{!}{\includegraphics{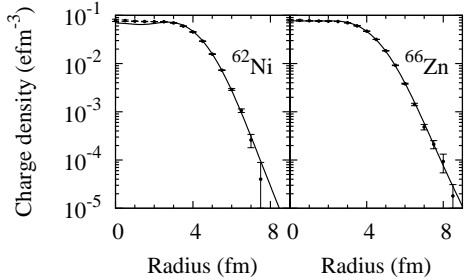}}
\caption{\label{chden} Calculated charge density in $^{62}$Ni and $^{66}$Zn (solid lines)
compared with experimental measurements (filled circles).
}
\end{figure}

The charge density has been obtained from the point proton
density $\rho_p$ by convoluting with a Gaussian form factor
to account the finite size of the proton.
 We plot in Fig. \ref{chden} the calculated charge density for
$^{62}$Ni and $^{66}$Zn as representatives of our results. One can see that the theoretical and experimental values agree very
well, particularly at larger radii values, which is the region expected to
contribute to the optical potential at low energy. Other nuclei
also show similar agreement.

\subsection{Decay}

Life time values have been calculated for different types of decays,
{\em i.e.} proton, alpha and cluster radioactivity. We employ the 
super-asymmetric fission model (SAFM). The decay product is assumed to tunnel through
the barrier created by the repulsive Coulomb (and the centrifugal) potential and attractive nuclear 
potential, constructed by folding.
The probability of barrier penetration has been calculated in the WKB 
approximation. The assault frequency is obtained from
the zero-point vibration energy\cite{gambhir} that, in turn, has been calculated from the
Q-values. 
Unlike some other works, we have not normalized the optical potential, 
and have used the values that were obtained from nuclear matter calculations,
for the decay studies.

Many calculations exist for proton radioactivity\cite{prot}.
Our results for the proton radioactivity half life calculation, in nuclei
where such decay has been observed from low energy states, are tabulated in
Table \ref{life1}. For comparison, we have also presented the results for 
the DDM3Y interaction of Basu \etal\cite{oth}, who have used a simple 
phenomenological form of the densities. We have also shown the 
uncertainties in the calculated half life values corresponding to the errors in the measured Q-values within parentheses. Our results deviate from experimental measurements in $^{147}$Tm, $^{150}$Lu and $^{156}$Ta, and more prominently in 
$^{185}$Bi and $^{177}$Tl$^*$, the last two results being off by an order of 
magnitude. One can show that the last two discrepancies can be explained as 
the effect of configuration mixing\cite{prot}.

\begin{table}
\caption{
Proton decay half lives (T) for spherical proton
emitters with even neutron number. Results are for the DDM3Y interaction.
The
angular momentum of the proton involved is given by $l$. 
\label{life1}}
\begin{tabular}{|l|c|l|l|l|}\hline
Nucleus & $l$  &\multicolumn{3}{c|}{log$_{10}$T(s)}\\\cline{3-5}
        &($\hbar$)     &\multicolumn{1}{c|}{Exp.}  &\multicolumn{1}{c|}{Present} & Ref\cite{oth}\\\hline
$^{105}$Sb  &  2 &   ~~2.049$_{-0.067}^{+0.058}$ & ~~2.27(46) & ~~1.97(46)   \\
$^{109}$I   &  2 & $-3.987_{-0.022}^{+0.020}$ & $-4.03(4)$ &  $-4.25$\\
$^{113}$Cs  &  2 & $-4.777_{-0.019}^{+0.018}$ & $-5.34(4)$ &  $-5.53$\\
$^{145}$Tm  &  5 & $-5.4097_{-0.146}^{+0.109}$  & $-5.20(6)$& $ -5.14(6)$ \\
$^{147}$Tm  &  5 &  ~~0.591$_{-0.175}^{+0.125}$ & ~~0.98(4) &  ~~0.98(4) \\
$^{147}$Tm* &  2 & $-3.444_{-0.051}^{+0.046}$ & $-3.26(6)$ &    $-3.39(5)$ \\
$^{151}$Lu  &  5 & $-0.896_{-0.012}^{+0.011}$ & $-0.65(3)$ &   $-0.67(3)$ \\
$^{151}$Lu* &  2 & $-4.796_{-0.027}^{+0.026}$ & $-4.72(10)$ &  $-4.88(9)$ \\
$^{155}$Ta  &  5 & $-4.921_{-0.125}^{+0.125}$ & $-4.67(6)$ &   $-4.65(6)$ \\
$^{157}$Ta  &  0 & $-0.523_{-0.198}^{+0.135}$ & $-0.21(11)$ &  $-0.43(11)$ \\
$^{161}$Re  &  0 & $-3.432_{-0.049}^{+0.045}$ & $-3.28(7)$ &   $-3.46(7)$ \\
$^{161}$Re* &  5  & $-0.488_{-0.065}^{+0.056}$ & $-0.57(7)$ &  $-0.60(7)$ \\
$^{165}$Ir* &  5  & $-3.469_{-0.100}^{+0.082}$ & $-3.52(5)$ &   $-3.51(5)$ \\
$^{171}$Au  &  0   & $-4.770_{-0.151}^{+0.185}$ & $-4.84(15)$ &  $-5.02(15)$ \\
$^{171}$Au* &  5  & $-2.654_{-0.060}^{+0.054}$& $-3.03(4)$ &    $-3.03(4)$ \\
$^{177}$Tl  &  0   & $-1.174_{-0.349}^{+0.191}$ & $-1.17(25)$ &  $-1.36(25)$ \\
$^{177}$Tl* &  5   & $-3.347_{-0.122}^{+0.095}$ & $-4.52(5)$ &  $-4.49(6)$ \\
$^{185}$Bi  &  0   & $-4.229_{-0.081}^{+0.068}$ & $-5.33(13)$ & $-5.44(13)$ \\
$^{112}$Cs  &  2 & $-3.301_{-0.097}^{+0.079}$    & $-2.93(11)$ & $-3.13$\\
$^{150}$Lu  &  5 & $-1.180_{-0.064}^{+0.055}$    & $-0.59(4)$ &$-0.58(4)$ \\
$^{150}$Lu* &  2 & $-4.523_{-0.301}^{+0.620}$    & $-4.24(15)$ & $-4.38(15)$ \\
$^{156}$Ta  &  2 & $-0.620_{-0.101}^{+0.082}$    & $-0.22(7)$ &$-0.38(7)$ \\
$^{156}$Ta* &  5 & $0.949_{-0.129}^{+0.100}$   & $~~1.66(10)$ & $~~1.66(10)$\\
$^{160}$Re  &  2 & $-3.046_{-0.056}^{+0.075}$   & $-2.86(6)$ &$-3.00(6)$ \\
$^{164}$Ir  &  5 & $-3.959_{-0.139}^{+0.190}$   & $-3.95(5)$ &$-3.92(5)$ \\
$^{166}$Ir  &  2 & $-0.824_{-0.273}^{+0.166}$   & $-0.96(10)$ & $-1.11(10)$ \\
$^{166}$Ir* &  5 & $-0.076_{-0.176}^{+0.125}$    & $~~0.22(8)$ & $~0.21(8)$ \\
\hline
\end{tabular}
\end{table}

It was often suggested that the half life values for proton radioactivity
of $^{109}$I and $^{112,113}$Cs could not be reproduced without the
inclusion of deformation effects. For example, a deformation
of the order of $\beta\sim 0.05-0.15$ was judged to be essential to explain 
the observed data.
However, our calculation reproduces the data for $^{109}$I with considerable
accuracy. In the deformed calculations, it is usually assumed that the
deformation of the parent and the daughter nuclei are identical. The daughter
in this particular case is $^{108}$Te. Te nuclei are well known vibrational
nuclei with very small deformation. One possibility may be that the
deficiency of the Woods Saxon potential far away from the stability valley
was responsible
for the failure of the earlier calculations. We also stress that our results are
nearly identical for both the NN interactions. The results for
$^{113}$Cs and to some extent $^{112}$Cs are not reproduced so well, which
may be an effect of deformation. However, in none of them do
we have an order of magnitude disagreement between theory and experiment as
obtained in certain other calculations.


Alpha decay have been studied in Refs. \cite{alpha1,alpha2,alpha3}.
For emission of complex products, there is another factor, the spectroscopic 
factor, introduced to incorporate the preformation probability. It contains 
the nuclear structure effects, and may be thought as the overlap between the 
actual ground state configuration of the parent and the configuration described 
by the complex decay product coupled to the ground state of the daughter. 
Obviously, it is expected to be much less than unity as there are contributions 
from many other configurations other than the one mentioned above. 

\begin{figure}[htb]
\resizebox{\columnwidth}{!}{\includegraphics{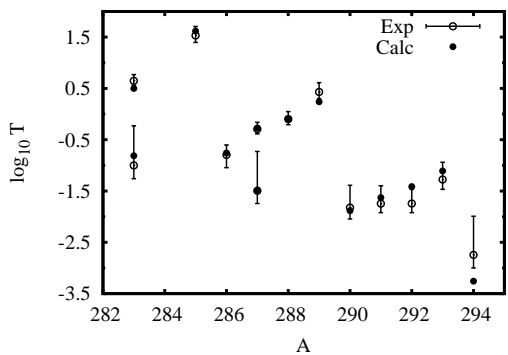}}
\caption{Calculated and experimental half life values in superheavy nuclei
\label{figP1}}
\end{figure}

In a small mass region, we do not expect the spectroscopic
factor to vary to any large extent. For example, in the superheavy nuclei
presented in Fig. \ref{figP1}, we  have taken a constant value
$1.4\times10^{-2}$ for
all the decays from a fit of the half life values\cite{alpha1}. In some other calculations,
the spectroscopic factors have been calculated from theory or phenomenological
formulas have been obtained for them\cite{alpha2}. 

\begin{figure}[htb]
\resizebox{\columnwidth}{!}{\includegraphics{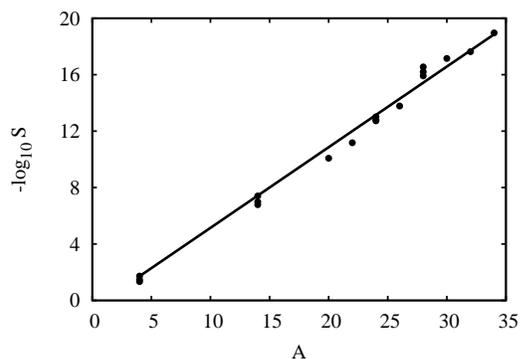}}
\caption{Negative of logarithm of spectroscopic factors ($S$) as a function of
cluster mass number $A$ for even-even parents and daughters.
\label{cl1}}
\end{figure}

The above method may be extended to study emission of complex clusters
\cite{Amiya,cluster1}.
Basu \cite{basu} has studied the nuclear cluster radioactivity
also in the framework of SAFM using a phenomenological density and the
realistic M3Y interaction. Bhagwat and Gambhir \cite{Amiya} have used densities
from RMF calculations. It has been suggested\cite{cluster} that in the case of decay of heavy
clusters,
the spectroscopic factor may scale as 

\be S=(S_\alpha)^{(A-1)/3}\ee
where $A$ is the mass of the heavy cluster and $S_\alpha$ is the
spectroscopic factor for the $\alpha$-decay. Thus a plot of  $\log_{10} S$ against
$A$ should be a straight line. In Fig. \ref{cl1}, we have plotted the
negative of $\log_{10} S$ for the decays where both the parent and the daughter are
even-even nuclei against the mass number of the cluster and plotted
a best fit line. One can see that the points fall nearly on a straight
line.


\subsection{ Elastic scattering}

\begin{figure*}[htb]
\begin{center}
\vskip -1.5cm
\resizebox{8cm}{!}{\includegraphics{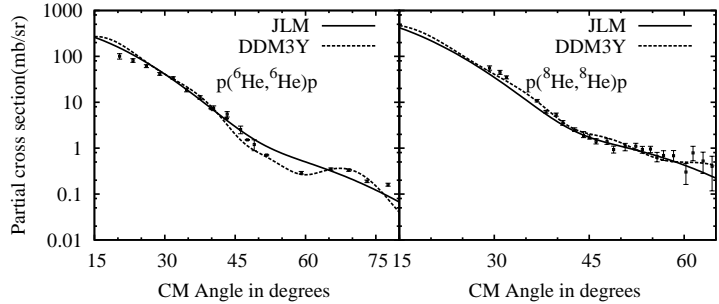}}
\caption{Partial cross section for elastic proton scattering
in inverse kinematics. The projectile energies of $^{6}$He and $^8$He are
71 MeV/A and 72 MeV/A, respectively.\label{Hescat}}
\end{center}
\end{figure*}

Elastic proton scattering in inverse kinematics provides a test for the 
calculated densities, particularly in absence of electron scattering data. 
Density information is not available in exotic neutron rich nuclei in very 
light mass region.
Elastic scattering cross sections for scattering of these nuclei from proton 
target have been calculated\cite{CBe} with the optical model potential 
generated in the semi-microscopic approach. In Fig. \ref{Hescat}, the 
measured partial cross sections for proton
scattering of $^{6,8}$He in inverse kinematics have been compared with theory.
The real and the imaginary parts of the potential have been chosen as 0.8 times and 0.2 times
the folded value for the DDM3Y potential. One can see that the experimental 
values are nicely reproduced.

\subsection{Low energy proton reactions of astrophysical significance}

Capture and charge exchange reactions at very low energy play a very important 
role in nucleosynthesis. Particularly, rapid proton capture ($rp$) process in 
explosive nucleosynthesis is a basic ingredient in driving the abundance along 
the $N=Z$ line\cite{book}. As this process has to overcome a large Coulomb 
barrier, it can occur only at a higher temperature range. For example, X-ray 
bursts provide large fluxes of protons at peak temperatures around 1-2 GK and 
are expected to play a significant role in the creation of nuclei up to mass 
110. 

The $rp$-process proceeds along the $N=Z$ line in mass 60-80 region. In nature,
the important proton capture reactions may involve certain nuclei as
targets which are not available to us. Hence, experimental information is
difficult, if not impossible, to obtain, at least in near future. In such a
situation, one has to rely on theory for the reaction rates.
Cross sections were calculated for proton capture reactions in mass 
60-80 region using semi-microscopic optical potentials in the local density
approximation with phenomenological density prescriptions. However, far from
the stability valley, these prescriptions may not represent the actual
densities very well, leading to considerable uncertainties in the reaction
rates. Very often, the reactions rates are varied by a large factor to
study their effects. For example, Schatz\cite{old} varied the rates of certain
reactions by a factor of one hundred. Obviously, this makes the results
uncertain to some extent and affects the final abundance.

A fully microscopic calculation may be used to estimate the rates to reduce
the above uncertainty. A consistent framework for calculation may be
constructed based on microscopic densities and may be extended to unknown mass 
regions with confidence. In the present work, we have tried to calculate
the reaction rates taking densities from a purely microscopic model, {\em i.e.} RMF.
We have already mentioned that RMF is particularly suitable to describe nuclei
far away from the stability valley where experimental knowledge is scarce.
A semi-microscopic optical potential obtained by folding an appropriate
microscopic NN interaction is expected to be more accurate and may do away
with the necessity of any arbitrary variation in the reaction rates.

However, even in a semi-microscopic optical potential, there often remain 
certain parameters that can be fixed only after comparison
with experiment. We have compared the results for a number of $(p,\gamma)$ and 
$(p,n)$ reactions in 
mass region $A=60-80$ for which experimental cross sections are available. This 
has helped us in determining a set of parameters for this mass region.

\begin{figure*}[hbt]
\vskip -4.5cm
\resizebox{!}{!}{\includegraphics{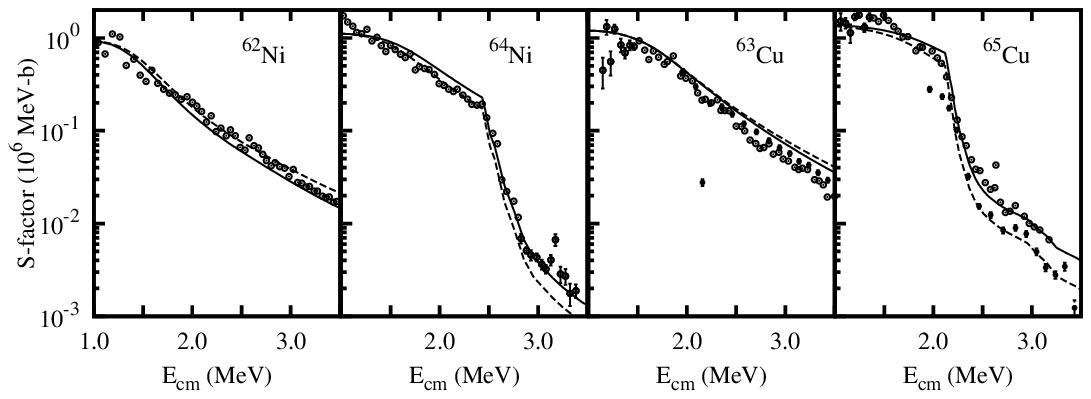}}
\caption{\label{nicusf}S-factors extracted from experimental measurements
compared with theory for $^{62,64}$Ni and $^{63,65}$Cu targets. Solid and dashed lines 
indicate respectively the results of the HF+BCS and HFB approaches for level 
density and E1 gamma strength. 
}
\vskip -4cm
\resizebox{!}{!}{\includegraphics{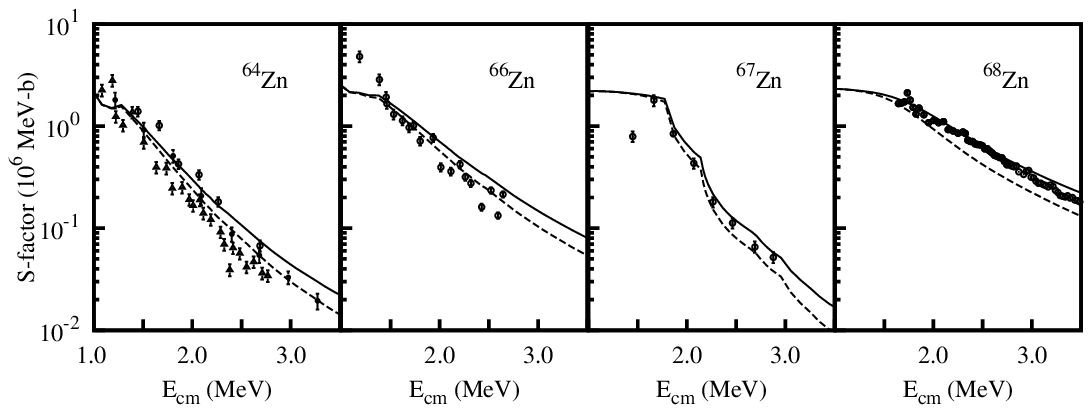}}
\caption{\label{znsf}S-factors extracted from experimental measurements
compared with theory for $^{64,66-68}$Zn. See caption of Fig. \ref{nicusf}
for details.}
\end{figure*}

The reaction calculations have been performed with the computer code 
TALYS 1.2\cite{talys} assuming spherical
symmetry for the target nuclei.
The real part of the potential has been obtained by normalizing the folded
DDM3Y potential by a factor of 0.7, while the imaginary part, by a factor of
0.1, so as to explain the S-factors obtained in the above experiments.
 We have
employed the full Hauser-Feshbach calculation with transmission coefficients
averaged over total angular momentum values and with corrections due to width
fluctuations. 
The gamma
ray strength has been calculated in the HFB  and HF+BCS
model.

In Figs. \ref{nicusf} and \ref{znsf}, we plot the results of our calculations for 
S-factors and compare them with experimental values for ($p,\ga$) reactions
for various targets in mass 60-70 region. 
The energy relevant to the $rp$-process in this mass region lies between
1.1 to 3.5 MeV. As the cross-section varies very rapidly at such low energy,
a comparison between theory and experiment is rather difficult. The usual
practice in low energy nuclear reaction is to compare another key observable,
{\em viz.} the S-factor. It is given by
\begin{equation}
S(E)=E\sigma(E)e^{2\pi\eta}
\end{equation}
where $E$ is the energy in centre of mass frame in keV, $\sigma(E)$ indicates 
reaction cross-section in barn and $\eta$ is the Sommerfeld parameter with
\begin{equation}
 2\pi\eta=31.29 Z_{p}Z_{t}\sqrt{\frac{\mu}{E}}
\end{equation}
Here, $Z_{p}$ and $Z_{t}$ are the charge numbers of the projectile and the
target, respectively and
$\mu$ is the reduced mass (in amu). The quantity S-factor varies much more 
slowly than reaction
cross-sections as the exponential energy dependence of cross-section
is not present in it. For this reason, we calculate this quantity
and compare it with experimentally extracted values.

\begin{figure}[hbt]
\vskip -4cm
\resizebox{12cm}{!}{\includegraphics{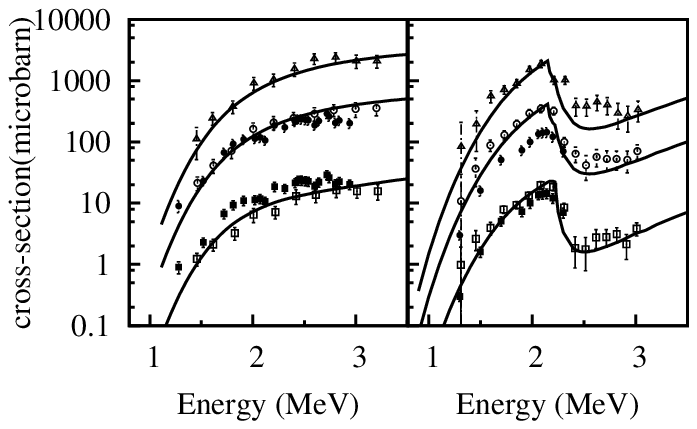}}
\caption{Partial cross sections for ($p,\ga$) reactions to the low-lying 
states in Zn for $^{63}$Cu (left panel) and $^{65}$Cu (right panel)
targets. 
Squares, circles and triangles represent data for transition to the ground 
state, the first excited state (multiplied 
by 10) and the second excited state (multiplied by
100), respectively. \label{partial}}
\end{figure}

In Fig. \ref{partial}, we plot the 
results for cross sections to populate the ground state and the first two 
excited states for $^{63,65}$Cu($p,\ga$) 
reactions. Results for proton capture by other 
nuclei as well as charge exchange reactions have also been calculated in this mass region and found to
agree reasonably well with experimental observations\cite{gambhir2,NiCu,wp,mass80,mass}.

Once the parameters have been fixed, we employ them to calculate the rates of
a number of astrophysically important proton reactions. Some very important
$N=Z$ nuclei, which have the highest abundance in an equilibrium in a chain,
are termed as waiting points\cite{book} for the chain. These nuclei have
negative or small positive Q-values for proton capture. An equilibrium between
the $(p,\gamma)/(\gamma,p)$ processes is established and the $rp$ process may
have to wait for beta-decay or $\alpha$-capture to proceed to heavier nuclei.
Certain $N=Z$ waiting point nuclei with $A<80$, {\em viz.} $^{64}$Zn,
$^{68}$Se, $^{72}$Kr, and $^{76}$Sr have long half lives, their total lifetime
being large compared to the time scale of typical X-ray bursts (10-100 sec).
Thus, they may produce a bottleneck in the
$rp$-process that would slow down the rate of hydrogen burning and necessitate
extended burst tails unless two proton capture can reduce these half lives
and bridge the waiting points.
X-ray burst model calculations are therefore particularly sensitive to the
rates of proton capture for these nuclei. We have used the microscopic approach,
outlined in the present work, to calculate the rates with an aim to study the
bridging of the waiting point nuclei.

For calculation of proton capture at waiting points, a small network, which 
includes the following processes, has been employed. The waiting point nucleus 
with $Z=N$, which acts as a seed, may capture a proton. The resulting nucleus, 
with $Z=N+1$, may either capture another proton or undergo photodisintegration
emitting a proton to go back to the seed nucleus. The nucleus with $Z=N+2$ may
also undergo photodisintegration. In addition, all the three nuclei mentioned
above may undergo $\beta$-decay. The photodisintegration rates at different
temperatures have been calculated from the proton capture rates using the
principle of detailed balance. The density has been taken as $10^6$ gm/cm$^3$
unless otherwise mentioned. The proton fraction has been assumed to be 0.7.

\begin{figure}[hbt]
\resizebox{\columnwidth}{!}{\includegraphics{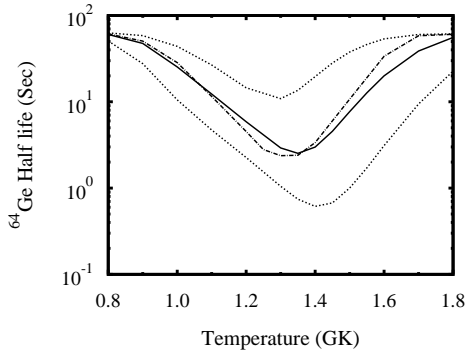}}
\caption{\label{ge64} Effective half life values of $^{64}$Ge as a function of 
temperature. The solid line represents the results of our calculation while
the dotted lines mark the two extremes for the errors in the Q-values of the
reactions involved. The dash dotted line shows the results obtained using the 
rates from \cite{astro}. }
\end{figure}

Fig. \ref{ge64} shows the change in the effective half life of $^{64}$Ge in 
explosive hydrogen rich environment. For
comparison, we have also plotted the results calculated from the rates in
Rauscher 
\etal\cite{astro} by dash-dotted lines.
The effects of the uncertainties on the half life values in the Q-values have
been indicated in the figures by dotted lines. 
The half life decreases and possibly goes  to a value substantially less
than ten seconds, the minimum duration of an X-ray burst. However, one sees
that the uncertainty in mass measurement prevents one from reaching any firm conclusion.
Depending on the actual value of the masses, it may even be possible that a
burst of the order of ten seconds cannot bridge this waiting point effectively.

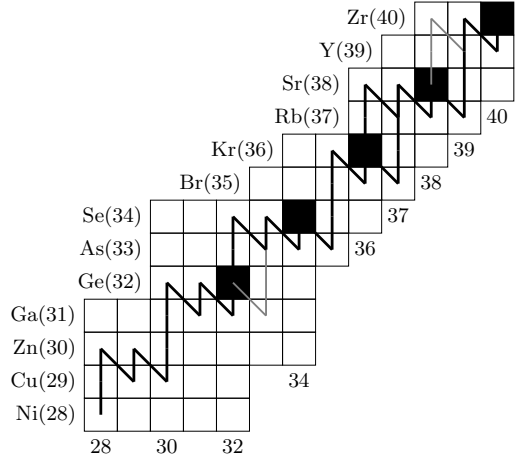
\begin{figure}[hbt]
\resizebox{\columnwidth}{!}{
\begin{tikzpicture}
\node (x1) at (0,-3)  {};
\node (x2) at (6,-3)  {};

  \node (1) at (.5,0)  [draw,inner sep=7.1pt,label=left:Ni(28),label=below :28] {$ $  };
    \node (2) at (.5,.5) [draw,inner sep=7.1pt,label=left:Cu(29)] {$ $};
    \node (3) at (.5,1) [draw,inner sep=7.1pt,label=left:Zn(30)] {$ $};
    \node (4) at (.5,1.5) [draw,inner sep=7.1pt,label=left:Ga(31)] {$ $};
    \node (5) at (1,.5) [draw,inner sep=7.1pt]{$ $};
    \node (6) at (1.,1) [draw,inner sep=7.1pt]{$ $};
 \node (7) at (1.,1.5) [draw,inner sep=7.1pt]   {$ $ };
 \node (8) at (1.5,.5) [draw,inner sep=7.1pt]    {$ $};
 \node (9) at (1.5,1) [draw,inner sep=7.1pt]    {$ $};
 \node (10) at (1.5,1.5) [draw,inner sep=7.1pt]   {$ $};
\node (11) at (1.5,2.) [draw,inner sep=7.1pt,label=left:Ge(32)]    {$ $};
 \node (12) at (2.,.5) [draw,inner sep=7.1pt] {$ $};
 \node (13) at (2.0,1) [draw,inner sep=7.1pt] {$ $};
 \node (14) at (2.,1.5) [draw,inner sep=7.1pt] {$ $};
\node (15) at (2.,2.) [draw,inner sep=7.1pt]  {$ $};
 \node (16) at (2.5,1.5) [draw,inner sep=7.1pt]   {$ $};
\node (17) at (2.5,2.) [fill,draw,inner sep=7.1pt]    {$ $};
 \node (18) at (2.5,2.5) [draw,inner sep=7.1pt]   {$ $};
\node (19) at (2.5,3.) [draw,inner sep=7.1pt]    {$ $};
 \node (20) at (3.,1.5) [draw,inner sep=7.1pt] {$ $};
 \node (21) at (1.0,0) [draw,inner sep=7.1pt]  {$ $};
 \node (22) at (1.5,0) [draw,inner sep=7.1pt,label=below :30]   {$ $};
 \node (23) at (2.0,0) [draw,inner sep=7.1pt] {$ $};
 \node (24) at (2.5,0) [draw,inner sep=7.1pt,label=below :32]   {$ $};
 \node (25) at (2.5,.5) [draw,inner sep=7.1pt]   {$ $};
 \node (26) at (2.5,1) [draw,inner sep=7.1pt]   {$ $};
\node (27) at (1.5,2.5) [draw,inner sep=7.1pt,label=left:As(33)] {$ $};
\node (28) at (1.5,3.) [draw,inner sep=7.1pt,label=left:Se(34)] {$ $};
\node (29) at (2.,2.5) [draw,inner sep=7.1pt] {$ $};
\node (30) at (2.,3.) [draw,inner sep=7.1pt] {$ $};
\node (31) at (3.,2.) [draw,inner sep=7.1pt] {$ $};
\node (32) at (3.,2.5) [draw,inner sep=7.1pt] {$ $};
\node (33) at (3.,3.) [draw,inner sep=7.1pt] {$ $};
\node (34) at (3.,3.5) [draw,inner sep=7.1pt,label=left:Br(35)] {$ $};
 \node (35) at (3.5,1) [draw,inner sep=7.1pt,label=below :34] {$ $};
 \node (36) at (3.5,1.5) [draw,inner sep=7.1pt] {$ $};
 \node (37) at (3.5,2.) [draw,inner sep=7.1pt] {$ $};
\node (38) at (3.5,2.5) [draw,inner sep=7.1pt] {$ $};
\node (39) at (3.5,3) [fill,draw,inner sep=7.1pt] {$ $};
\node (40) at (3.5,3.5) [draw,inner sep=7.1pt] {$ $};
\node (41) at (3.5,4.) [draw,inner sep=7.1pt,label=left:Kr(36)] {$ $};
\node (42) at (4.,2.5) [draw,inner sep=7.1pt] {$ $};
\node (43) at (4.,3.) [draw,inner sep=7.1pt] {$ $};
\node (44) at (4.0,3.5) [draw,inner sep=7.1pt] {$ $};
\node (45) at (4.0,4.) [draw,inner sep=7.1pt] {$ $};
\node (46) at (4.5,3.) [draw,inner sep=7.1pt,label=below:36] {$ $ };
\node (47) at (4.5,3.5) [draw,inner sep=7.1pt] {$ $ };
\node (48) at (4.5,4.) [fill,draw,inner sep=7.1pt] {$ $};
\node (49) at (4.5,4.5) [draw,inner sep=7.1pt,label=left:Rb(37)] {$ $ };
\node (50) at (4.5,5.) [draw,inner sep=7.1pt,label=left:Sr(38)] {$ $ };
\node (51) at (5.,3.5) [draw,inner sep=7.1pt,label=below:37] {$ $ };
\node (52) at (5,4.) [draw,inner sep=7.15pt] {$ $  };
\node (53) at (5.,4.5) [draw,inner sep=7.1pt] {$ $ };
\node (54) at (5.,5.) [draw,inner sep=7.1pt] {$   $ };
\node (55) at (5.,5.5) [draw,inner sep=7.1pt,label=left:Y(39)] {$  $ };
\node (56) at (5.5,4.) [draw,inner sep=7.1pt,label=below:38] {$ $ };
\node (57) at (5.5,4.5) [draw,inner sep=7.1pt] {$  $ };
\node (58) at (5.5,5) [fill,draw,inner sep=7.1pt] {$ $ };
\node (59) at (5.5,5.5) [draw,inner sep=7.1pt] {$ $ };
\node (60) at (5.5,6.) [draw,inner sep=7.1pt,label=left:Zr(40)] {$ $ };
\node (61) at (6.,4.5) [draw,inner sep=7.1pt,label=below:39] {$ $ };
\node (62) at (6.,5.) [draw,inner sep=7.1pt] {$ $ };
\node (63) at (6.,5.5) [draw,inner sep=7.1pt] {$ $ };
\node (64) at (6.,6.) [draw,inner sep=7.1pt] {$ $ };
\node (66) at (6.5,5.) [draw,inner sep=7.1pt,label=below:40] {$ $ };
\node (67) at (6.5,5.5) [draw,inner sep=7.1pt] {$ $ };
\node (68) at (6.5,6.) [fill,draw,inner sep=7.1pt] {$ $ };
\node (71) at (3.,1.) [draw,inner sep=7.1pt] {$ $ };

\draw [very thick]  (.5,0) to (.5,.5); 
\draw [very thick]  (.5,.5) to (.5,1.); 
\draw [very thick]  (.5,1.) to (1.0,.5); 
\draw [very thick]  (1.0,.5) to (1.0,1.); 
\draw [very thick]  (1.0,1.) to (1.5,.5); 
\draw [very thick]  (1.5,.5) to (1.5,1.); 
\draw [very thick]  (1.5,1) to (1.5,1.5); 
\draw [very thick]  (1.5,1.5) to (1.5,2.); 
\draw [very thick]  (1.5,2.) to (2.,1.5);
\draw [very thick]  (2.,1.5) to (2.,2.);
\draw [very thick]  (2.,2.) to (2.5,1.5); 
\draw [very thick]  (2.5,1.5) to (2.5,2.);
\draw [very thick]  (2.5,2.) to (2.5,2.5);
\draw [thick, gray]  (2.5,2.) to (3.0,1.5);
\draw [thick, gray]  (3.0,1.5) to (3.0,2.5);
\draw [very thick]  (2.5,2.5) to (2.5,3.);
\draw [very thick]  (2.5,3.) to (3.,2.5);
\draw [very thick]  (3.,2.5) to (3.,3.);
\draw [very thick]  (3.,3.) to (3.5,2.5);
\draw [very thick]  (3.5,2.5) to (3.5,3.);
\draw [very thick]  (4.,3.5) to (4.,4.);
\draw [very thick]  (4.,4.) to (4.5,3.5);
\draw [very thick]  (4.5,3.5) to (4.5,4.);
\draw [very thick]  (4.5,4.) to (4.5,4.5);
\draw [very thick]  (4.5,4.5) to (4.5,5.);
\draw [very thick]  (4.5,5) to (5,4.5);
\draw [very thick]  (5.,4.5) to (5,5.);
\draw [very thick]  (5,5.) to (5.5,4.5);
\draw [very thick] (5.5,4.5) to (5.5,5.);
\draw [very thick] (5.5,5) to (6,4.5);
\draw [very thick] (6,4.5) to (6,5.5);
\draw [thick, gray] (5.5,5) to (5.5,5.5);
\draw [thick, gray] (5.5,5.5) to (5.5,6.);
\draw [thick, gray] (5.5,6.) to (6,5.5);
\draw [very thick] (6,5.5) to (6.,6.);
\draw [very thick] (6.,6.) to (6.5,5.5);
\draw [very thick] (6.5,5.5) to (6.5,6.);
\draw [very thick] (4.5,4.) to (5,3.5);
\draw [very thick] (5,3.5) to (5,4.);
\draw [very thick] (5,4.) to (5,4.5);
\draw [very thick] (5,4.5) to (5,5.);
\draw [very thick] (3.5,3.) to (4.,2.5);
\draw [very thick] (4.,2.5) to (4.,3.);
\draw [very thick] (4.,3.) to (4.,3.5);

\end{tikzpicture}
}
\vskip -2cm
\caption{\label{fig:SRD} The rp-process path for 1.2 GK.}
\end{figure}

A larger network, starting from $^{56}$Ni seed and consisting of $(p,\gamma)$,
$(\gamma,p)$ reactions and $\beta$-decay, 
  has been employed to study the
the time evolution of abundance. In Fig. \ref{fig:SRD} we plot the path 
followed by the $rp$-process between $A=56$ and $A=80$ for $T=1.2$ GK. The 
dark boxes indicate the waiting point nuclei. 
The elements are indicated at the left of the diagram while the neutron numbers 
are shown at the bottom. The lines
indicate the path followed by nucleosynthesis. The paths through which
more than 10\% of the total flux flows are indicated by black lines while gray
lines show the corresponding paths for flux between 1\% - 10\%.
One can see that $^{64}$Ge is easily bridged by two proton capture at 1.2 GK.
However, we have found that at a higher temperature, {\em viz.} 1.5 GK, 
this waiting point delays the process so that 
the $\beta$-decay of  $^{64}$Ge contributes significantly. At the next 
waiting point, $^{68}$Se, the photodisintegration is sufficiently strong so 
that, independent of the temperature, $\beta$-decay is practically the only 
available path. This delays the nucleosynthesis significantly. At $^{72}$Kr, 
the inverse process predominates in  higher temperature driving the flux through
$\beta$-decay. Thus, here also, lower temperature helps nucleosynthesis speed 
up. The waiting point at $^{76}$Sr presents a different
picture where the path essentially does not depend on the temperature and
principally flows along decay. It is clear that the actual process is
significantly dependent on the model of the burst process where the temperature
and the proton fraction are functions of time.

\begin{figure}[hbt]
\hskip -0.5cm
\resizebox{6.5cm}{5cm}{\includegraphics{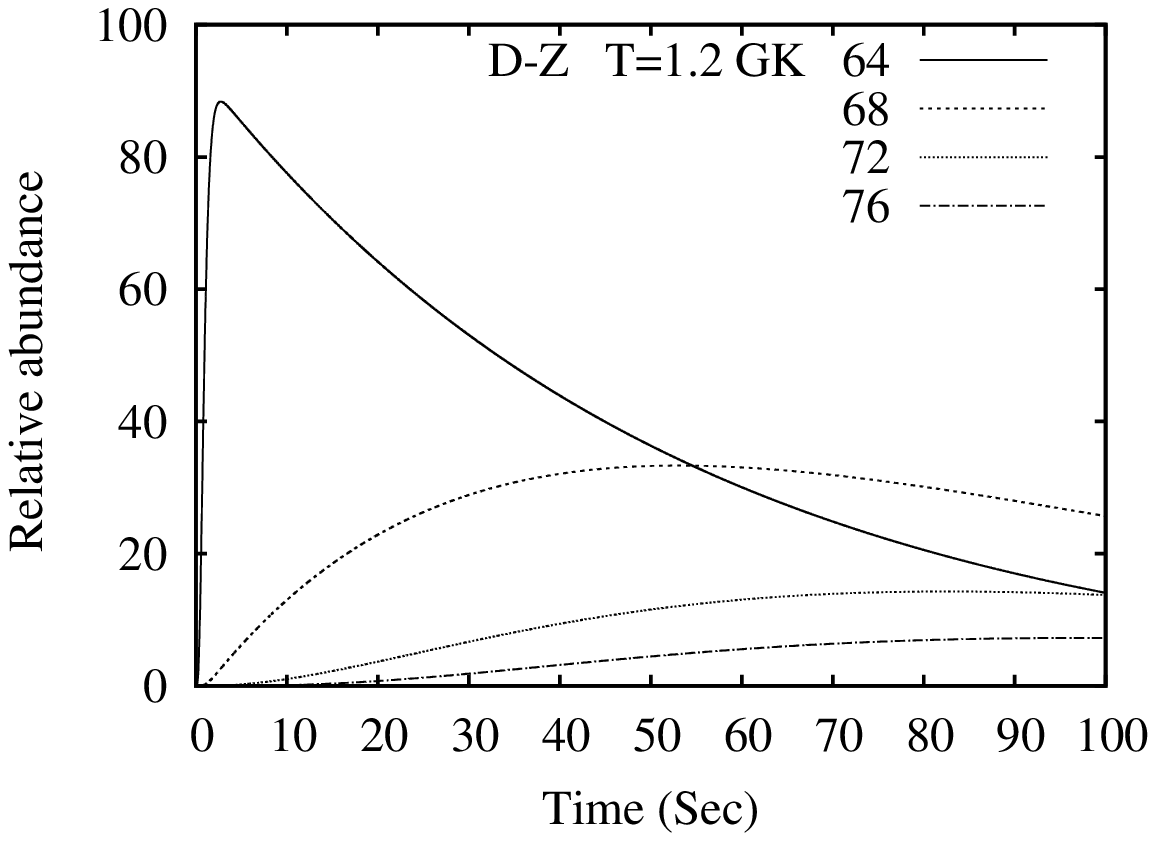}}\hskip -0.5cm 
\resizebox{6.635cm}{5cm}{\includegraphics{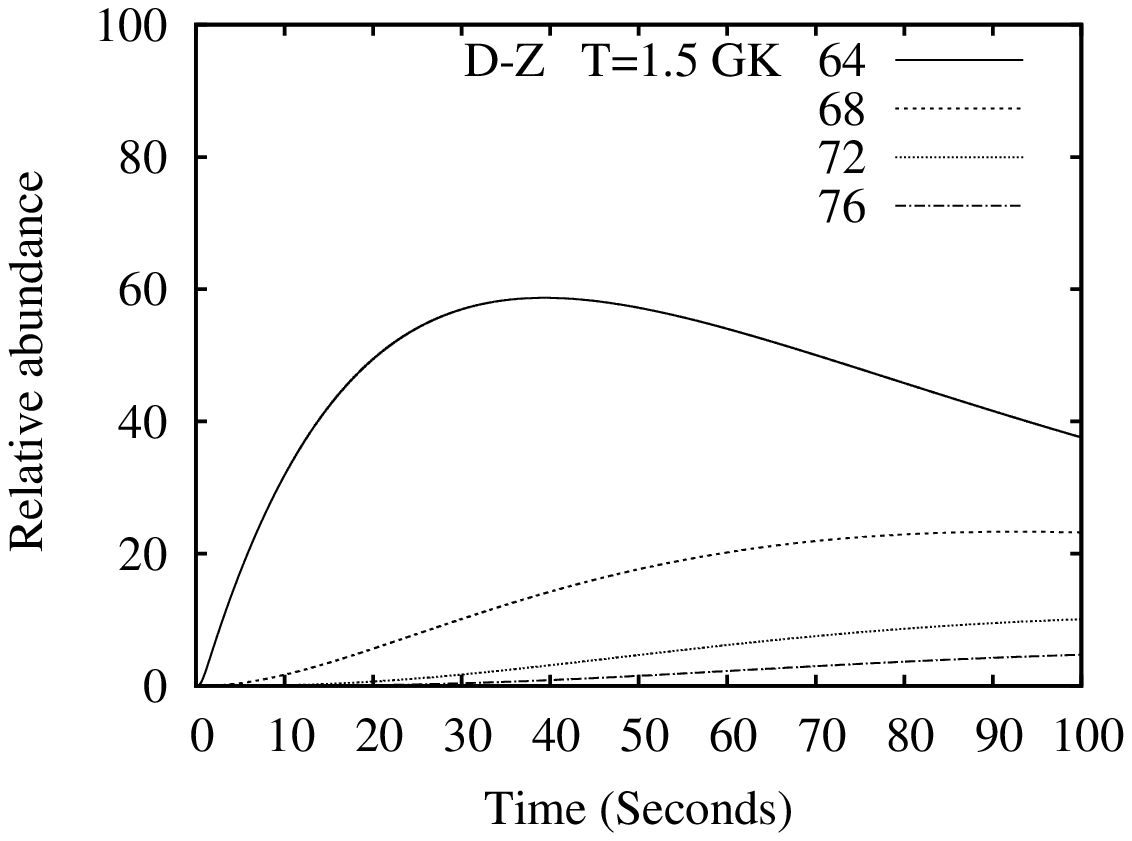}}\hskip -0.635cm 
\caption{Evolution of abundance of mass at the waiting points in explosive
proton-rich astrophysical environments. \label{net}}
\end{figure}

The abundance at two different temperatures are plotted as a function of time
in Fig. \ref{net}. 
We find that at 1.2 GK, at the end of 100 seconds, the population that reaches
$A=80$ or beyond is more than 1.5 times than the corresponding quantity at
the higher temperature of 1.5 GK. In both the cases, the population beyond
$A=76$ is significant. It should be noted that these numbers are strongly
dependent on the masses of the nuclei near the 
waiting point. Masses of these nuclei have either not been
measured, or measured very imprecisely. Thus one has to depend on theory. 
In fact, the final abundance at various masses change by a large factor as one 
uses estimates from different formulas. This conclusion is in agreement with the
variation in effective half life values seen in Fig. \ref{ge64}. 
In the 
present instance, the Duflo-Zuker formula\cite{DZ} has been used.

\section{Summary}

The techniques used to obtain semi-microscopic optical potentials from
theoretical densities using RMF and effective NN interactions have been 
described. Life time for emission of protons, neutrons and complex nuclei
from nuclei have been successfully calculated in the WKB approximation using 
the real part of the above potential. The potential has also been used to 
describe elastic scattering and proton reactions at low energy. Astrophysical 
implications of
the results of proton capture reactions have been discussed. Representative
results for above calculations have been presented. It is possible to conclude that
semi-microscopic optical potentials obtained in the folding model with RMF 
densities are very much successful in explaining nuclear decays and reactions
over the entire mass region.

\section*{Acknowledgments}
The work presented is actually a joint effort with contributions from
Madhubrata Bhattacharya, Chirashree Lahiri, Partha Roy 
Chowdhury, Abhijit Bhattacharyya and Subinit Roy. Financial assistances 
provided by the 
DAE-BRNS and the UGC sponsored
DRS Programme of the Department of Physics of the University of Calcutta are
gratefully acknowledged.



\end{document}